\documentclass[a4paper]{jpconf}
\usepackage[english]{babel}
\usepackage[utf8x]{inputenc}
\usepackage{amsmath}
\usepackage{graphicx}
\usepackage{booktabs}
\usepackage{multirow}
\usepackage{xspace}
\usepackage{color}
\usepackage[colorinlistoftodos]{todonotes}

\usepackage{capt-of}
\usepackage{booktabs}
\usepackage{varwidth}

\usepackage{blindtext}

\begin{document}

\title{Exploiting Apache Spark platform for CMS computing analytics}

\providecommand{\keywords}[1]{\textbf{\textit{CERN, CMS, HEP, Big Data, Hadoop, Spark, Analytics, Machine Learning, Zeppelin}} #1}

\author{Marco Meoni$^1$, Valentin Kuznetsov$^2$,
Luca Menichetti$^3$, Justinas Rumševičius$^4$
Tommaso Boccali$^5$ and 
Daniele Bonacorsi$^6$}
%for the CMS collaboration}

\address{
$^1$ University of Pisa \& INFN Pisa, Italy\\
$^2$ Cornell University, NY, U.S.A.\\
$^3$ CERN, Switzerland\\
$^4$ Vilnius University, Lithuania \\
$^5$ INFN Pisa, Italy\\
$^6$ University of Bologna and INFN, Italy}

\ead{marco.meoni@cern.ch,
vkuznet@gmail.com,
luca.menichetti@cern.ch,
justinas.rumsevicius@gmail.com,
tommaso.boccali@pi.infn.it,
daniele.bonacorsi@unibo.it}

%\date{28 August 2017}

\begin{abstract}
\boldmath 
The CERN IT provides a set of Hadoop clusters featuring more than 5 PBytes of raw storage with different open-source, user-level tools available for analytical purposes. The CMS experiment started collecting a large set of computing meta-data, e.g. dataset, file access logs, since 2015. These records represent a valuable, yet scarcely investigated, set of information that needs to be cleaned, categorized and analyzed. CMS can use this information to discover useful patterns and enhance the overall efficiency of the distributed data, improve CPU and site utilization as well as tasks completion time. Here we present evaluation of Apache Spark platform for CMS needs. We discuss two main use-cases CMS analytics and ML studies where efficient process billions of records stored on HDFS plays an important role. We demonstrate that both Scala and Python (PySpark) APIs can be successfully used to execute extremely I/O intensive queries and provide valuable data insight from collected meta-data.
\end{abstract}

%\keywords
{\bf keywords:} CERN, CMS, HEP, Big Data, Hadoop, Spark, Analytics, Machine Learning.

\section{Introduction} \label{Introduction}

The need for agile data analytics platforms is increasing in High Energy Physics (HEP). The current generation of HEP experiments has globally surpassed the Exa-Byte of storage with deployed computing resources exceeding one million computing cores. On top of this, computing infrastructures are highly distributed, and data transfer between sites is a major activity in the computing operations. 
Systems of such complexity require a careful monitoring in order to be maintained in a healthy state. At this scale processing TBytes of data on a daily basis become a real challenge. Big Data derived approaches have started to be introduced in order to get a better understanding of the computing operations and utilization of computing resources. Among others, the Elasticsearch, Hadoop Map-Reduce, and Spark are attractive solutions to handle large data sets.
CERN itself is moving most of its computing dashboards from a standard RDBMS oriented approach (current dashboards use Oracle as a back-end) towards Hadoop based solutions \cite{CERN-MONIT}.
The expected outcome of this transition is to efficiently process large data sets and aggregate information across distributed data providers.
Such studies are becoming more and more important due to the need of reducing computing resources and the associated cost with respect to  general increase of physics data collected by the LHC experiments. Funding agencies and review committees are, in fact, asking for increasingly optimized computing operations which can only benefit from a deeper understanding of computing resource utilization.

Since 2015, the Compact Muon Solenoid (CMS) experiment \cite{CMS} has stored large sets of computing logs (``meta-data'') about user accounting, jobs life-cycle, resources utilization, sites performance, software versions, and file access, on Hadoop file systems. The Apache Spark open-source cluster-computing framework has been evaluated as a valuable candidate to handle large amount of this meta-data stored on Hadoop file system (HDFS). Spark provides a scalable and high-performance data processing solution based on APIs offered in a variety of languages.

The benefit of using Spark for CMS experiment is two-fold. On one hand, it reduces the processing time on large datasets (DS) and manages effectively heterogeneous meta-data. On the other, it allows to perform Machine Learning (ML) studies and/or computing analytics over large DS from distributed data sources. Starting from unstructured raw data, the data preparation and analysis steps are completely defined in Hadoop infrastructure. This opens up a possibility to offload production RDBMS workload onto a Hadoop+Spark environment which is more suitable for large data aggregations and distributed joins.

In this paper we address two main use cases:
\begin{itemize}
\item CMS data popularity which provides various metrics about CMS user activities based on data available at CERN HDFS system. In particular, we discuss the aggregation pipeline from multiple data-streams and present useful information about user access patterns, such as number of accesses, CPU metrics versus data-tiers, sites, in visualization dashboards;
\item Intelligent data placement based on modeling (via ML) and predicting popular CMS datasets to be used in analysis jobs. This activity includes fast queries for reprocessing monitoring time intervals, file-to-DS aggregation and correlation, as well as the evaluation of predictive models.
\end{itemize}
The paper is organized as follows. Sect. \ref{RelatedWork} outlines the related works, with focus on efforts applied to Big Data analytics in HEP. Then we provide description of system setup in Sect. \ref{Setup} where we discuss technical parts of the data flow and underlying technologies. Then, we address the CMS data popularity use case in Sect. \ref{CMSPopularity}. Finally, we demonstrate how Hadoop and Spark platforms can be used to analyze large data sets in order to predict user behavior in Sect. \ref{Popularity} followed by discussion of using these predictions in sites caching strategies.

\section{Related Work} \label{RelatedWork}

After the introduction of Spark platform \cite{Spark} for the general public in 2010 and following adaptation of SQL \cite{SparkSQL} and ML \cite{SparkML} within Spark, many businesses realized its power within
the domain of data analytics. Small and large companies successfully applied this framework to different use cases which required processing large datasets. Up until now, the HEP uses GRID infrastructure to process individual jobs at distributed sites. In order to leverage Map-Reduce paradigm and distributed nature of Hadoop+Spark platform many aspects should be modified.  Among the others, changes to middle-ware in order to read underlying ROOT data-format, adaptation of data workflow scheduling and submission tools as well as overall design of data processing \cite{BigDataHEP}. Such effort is undergoing within recently funded projects such as DIANA-HEP \cite{DIANA-HEP}. The analytics use cases, e.g. extracting data insight from available meta-data (logs), is very well studied and adapted in a business world. Recently, HEP experiments realized that they can explore their meta-data to gain additional insight about user activities and apply this knowledge to improve their computing models \cite{CERN-MONIT}. In particular, exploiting Machine Learning techniques on computing meta-data is a hot topic in the HEP community at large \cite{HEP-WP}. The data popularity problem was studied by all major HEP experiments, e.g. in CMS \cite{CMSDataPopularity, CHEP2016}, LHCb \cite{LHCbDataPopularity}, and ATLAS \cite{ATLASDataPopularity}. It was found that ML approach can be useful addition to custom data placement solutions adopted by experiments. In this work, we extend previous CMS attempts \cite{CMSDataPopularity, CHEP2016} to use CMS meta-data in the context of dataset popularity and extend discussed techniques to Spark platform.

\section{System Setup and Implementation} \label{Setup}
The European Organization for Nuclear Research (CERN) has deployed two large Hadoop clusters featuring nearly 5 PBytes of raw storage. Currently, they consist of around 50 nodes equipped with 64 GBytes of RAM and 32 cores/node on a mix of Intel/AMD CPUs and CentOS7/SLC6 operating systems.

The Spark cluster is tuned and optimized to reach maximum scalability and parallelism. Every job is calibrated in favor of local processing within the same executor and reducing the data exchange between cluster nodes.
%Among the available options, we leverage \texttt{spark.core.max}, \texttt{spark.executor.cores} and \texttt{spark.dynamicAllocation.maxExecutors} cluster settings. 
%\lm{TODO: Luca explains what the 3 settings do}
The Spark paradigm consists of reading data once and processing them as much as possible in memory, minimizing data shuffling, network communication between executors, and disk I/O operations. In addition, together with the cluster resource manager (YARN in our case), Spark jobs can profit from dynamic allocation of their executors, automatically scaling in and out their number when necessary. This allows users to run the same job against folders of the same dataset without adapting different parameters according to their individual sizes.

% previous version: Since Spark is written in Scala, we use the latter one for data aggregation in order to avoid language translation while processing raw data. Moreover, its dataframe and RDD parallelism is available out-of-the-box along with in-memory and persistence API. On the other hand, Python is a language largely known in HEP, which makes its learning curve very quick. However PySpark, the Spark-Python API, is a wrapper around Java libraries that can be slower versus native libraries and may need to handle memory issue not known in Python. Certainly, applications' throughput highly depends on their structure and it is best practice to use dataframe operations instead of iterations.

% [Luca] the sentence
%%%  "Moreover, its dataframe and RDD parallelism is available out-of-the-box along with in-memory and persistence API." %%%
% implies that PySpark does not have those things, which is not true.
% I think it's dangerous to motivate why we use Scala mentioning that the python wrapper is slower. Spark developers have worked a lot to eliminate any performance issue with Python. In 2.0+ versions there are basically no differences.
%%% "may need to handle memory issue not known in Python" %%%
% memory issues come from Java: not to use Python because you have errors from java is not a valid argument IMO
% I would delete this part or write something like this:

The CERN Hadoop service provides scientists with a variety of tools and Big Data frameworks, such as Spark, Pig, Hive, which are available in different clusters. These are distinct but not isolated environments where data stored in one HDFS namespace can be accessed by jobs running indiscriminately in whatever cluster. Obviously jobs reading the local namespace will be faster, and if necessary data can be moved from one cluster to another using Map-Reduce jobs.
%(with the $distcp$ command).

CMS data is stored in daily folders in CSV, JSON, or Avro format. Spark 2.0 framework, via the SparkSQL context, is equipped  to read these formats. These components take as input an HDFS path and may require additional parameters depending on the kind of the format, e.g. a header of CSV format, a data schema or quotation symbols, and so on. All of them return a DataFrame used in forthcoming computations. 

Since Spark is written mainly in Scala and Java, we originally used the former language for work discussed in Sect. \ref{Popularity}. Later we found that the Scala programming language is not (yet) well adopted in HEP community. Therefore, we made significant effort to adopt our code base to PySpark framework, as explained in Sect. \ref{PySpark}. 
%Python is a language largely known in HEP, with a very convenient  learning curve; in most cases, CMS collaborators are already fluent in Python.
%However, DataFrame operations are used instead of iterations over classic data structures, hence not all the Python features can be employed in such context. 
%Parallelism is provided out-of-the-box via DataFrames and RDDs actions and transformations along with in-memory and persistence API, thus maintaining the same language while processing huge data volumes allows to overcome in advance any possible performance issue. 

In next Sections \ref{Scala} and \ref{PySpark}, we provide details of our experience with Scala and PySpark APIs. We mostly outline pros and cons of both languages based on our practice with them.

%The Spark platform was designed to process large amount of data efficiently by means of a simple programming interface. The APIs are divided into Unstructured and Structured ones; the former works with Resilient Distributed Datasets (RDDs), Accumulators, and Broadcast variables, while the latter gives access to DataFrames, Datasets and the SparkSQL Context. There are few languages which implement Spark APIs, the native solution is based on Scala programming language as well as there is a wrapper solution is available as a thin layer around Spark libraries. Even though usage of PySpark APIs is straightforward it is worthwhile to mention that the code based on PySpark APIs should be carefully crafted to run on Spark platform, i.e. you cannot invoke Python to run it and it requires to write a wrapper to submit such code for execution. Also we found that standard programming paradigms, such as for loops, sometimes become useless when writing spark code and should be replaced with PySpark counterparts to operate over Spark DataFrames.  

\subsection{Scala usage} \label{Scala}

\begin{figure}[ht]
\begin{center}
\includegraphics[width=4.5in]{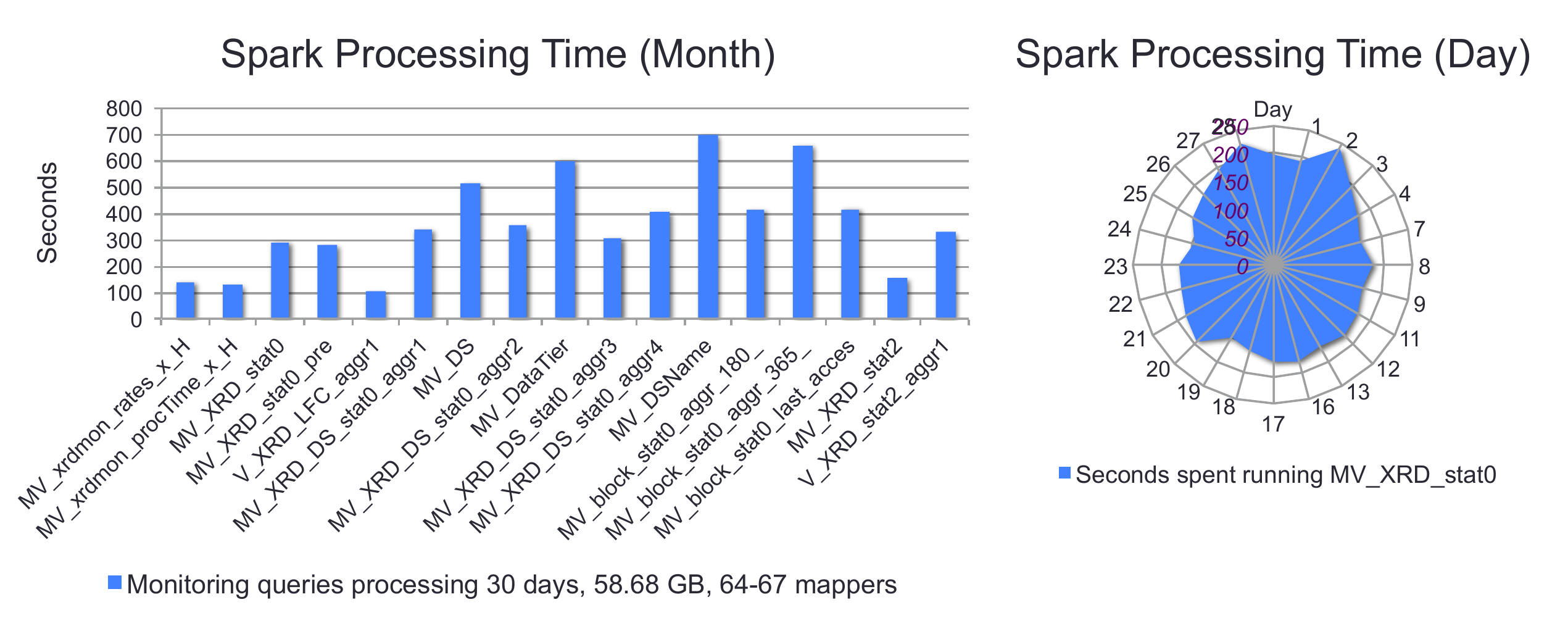}
\caption{Fast reprocessing of DS popularity monitoring query on Apache Spark. The left plot shows monthly processing time for individual Oracle DB views implemented in Spark. The plot on a right shows processing time of one Oracle DB view for daily records.}
\label{fig:fig-queries}
\end{center}
\end{figure}

Our original attempt to process large DS on HDFS was done in the Scala programming language and it reflects results described in Sect \ref{Popularity}. Here we would like to highlight our observation of using Scala for this task. 
Fig. \ref{fig:fig-queries} shows the processing time required for DS popularity monitoring queries. The first plot shows the amount of time, in seconds, that was needed by a number of monitoring queries. The names on the x-axis are borrowed from the corresponding materialized views (MV) implemented in the current Oracle-based monitoring infrastructure.
%Each query processes row file access logs according to different usage features and keywords in the dataset name syntax. 
The second plot shows the reprocessing time to reproduce the results of an Oracle DB view. We found that regardless of the Oracle view or selected time period the average processing time remains in a range of few minutes\footnote{
In order to exploit built-in distribution and persistence, we relied on \texttt{repartition()} and \texttt{persist()} API to split data uniformly across the computing nodes and cache them during the aggregation steps.}.
This makes Spark and Scala a very effective combination for reprocessing popularity queries on Hadoop. In fact, the same results take continuous running of incremental MV updates in Oracle back-end.

\subsection{PySpark usage} \label{PySpark}

%The Spark platform was designed to process large amount of data efficiently by means of a simple programming interface. The APIs are divided into Unstructured and Structured ones; the former works with Resilient Distributed Datasets (RDDs), Accumulators, and Broadcast variables, while the latter gives access to DataFrames, Datasets and the SparkSQL Context. There are few languages which implement Spark APIs, the native solution is based on Scala programming language as well as there is a wrapper solution is available as a thin layer around Spark libraries. Even though usage of PySpark APIs is straightforward it is worthwhile to mention that the code based on PySpark APIs should be carefully crafted to run on Spark platform, i.e. you cannot invoke Python to run it and it requires to write a wrapper to submit such code for execution. Also we found that standard programming paradigms, such as for loops, sometimes become useless when writing spark code and should be replaced with PySpark counterparts to operate over Spark DataFrames.  

The transition from Scala to PySpark was trivial. But
we found a few particularly interesting issues every programmer should know about when developing code using Python (PySpark) APIs.

Careful crafting of data is required to fit it into worker's node memory. For instance, most of the run-time errors we experienced were related to java.lang.OutOfMemoryError, Java heap, and GC errors issues. A typical example would be the temptation to use $df.collect()$ PySpark function to collect intermediate results. Instead, the code should perform series of operations over the data and either write them out to HDFS or run aggregation steps. Further, when the data are skewed not all workers obtain the right amount of workload. For instance, grouping or shuffling operations of non equally distributed data can cause
the slowness of entire job within executor. 
%The same job may run or fail depending on resource utilization and operations the code is doing to process large amount of data.
To take advantage of parallelism the code should not iterate over the containers, instead, optimized PySpark APIs should be used, e.g. $df.foreach$ in favor of $for\;item\;in\;df.collect()$ pattern or $df.treeReduce(lambda\;x,y:x+y)$ instead of local non cluster-distributed functions $np.sum(df.collect())$.

%Another example would be dropping list comprehensions in favor of nested calls to PySpark APIs, e.g grouping a data frame over some variable following by aggregation on certain attribute with final renaming of certain columns should be expressed as following:
%$$df.groupBy([XYZ]).agg({attr:"count"}).withColumnRenamed(X,Y)$$
%No global scope within container operations.

Users are advised to rely on PySpark functions to avoid slowness of built-in Python functions over DataFrame operations,
e.g. $pyspark.sql.functions.split$ should be used instead of $split$ Python built-in function.

%\item Different version of PySpark, e.g. 1.X and 2.X, provides different API set and coders should be careful to adopt their code to work properly in different PySpark releases;

%\item Lack of Python support for new Kafka streams which affects data streaming within Spark job.

To avoid these types of pitfalls we developed our own framework called CMSSpark \cite{CMSSpark} which provided all necessary tools for code submission to Spark platform, set up necessary Java libraries for data processing (e.g. support for CSV or Avro data formats) as well as set up proper run parameters optimized for CERN Spark cluster.

This framework significantly helped new developers to easily adopt their business logic into framework and quickly produce desired results. For instance, any user with access to the CERN Hadoop cluster can quickly launch distributed (Py)Spark job to process CMS data, focusing on what operation to run rather than struggling on how to properly submit YARN-based batch job.

\section{CMS Data Popularity} \label{CMSPopularity}

%In this section we discuss how Spark can offer a powerful platform for processing large data-volumes and providing insights into user activities. 
The common approach of data analytics is based on RDBMS solutions. At CERN, the Oracle DB is a de-facto standard for many bookkeeping and analytics services. Even though it works extremely well we see that the amount of information HEP experiments want to process is growing exponentially each year. We already found certain limitations on data processing pipelines within Oracle DB. For instance, the CMS most populated DBs, Data Bookkeeping System (DBS) and Data Location System (PhEDEx) \cite{CMSDataManagement}, already contain a few hundred GBytes of data. Merging data only from those two DBs becomes an issue. Even though, they reside in one physical DB the cross-joins among them caused large latencies on complex queries due to large data volume.
Moreover, growing number of collected meta-data about user jobs which resides outside of Oracle DB may provide an additional insight into data utilization by our infrastructure.
Spark platform provides a valuable solution to overcome these obstacles.

The CMS, among other HEP experiments, has joined a new trend of storing various meta-data to HDFS starting in 2015. This effort was driven by individual users and groups willing to investigate a new technology to analyze unstructured data, such as logs from user based jobs. By 2017 we already collected billions of records from various data providers.
Here we present a successful effort showing utilization of Hadoop+Spark platform for analytics purposes and modeling of DS popularity based on available meta-data.
We start our discussion with CMS analytics use case where significant portions of these data were used.

\subsection{Data Processing} \label{DataProcessing}

We used Spark jobs to acquire various information about user activities such as total number of accesses to dataset and processing time of analysis jobs at specific sites. We identified four data-streams (in the form of logs) representing this information: AAA \cite{AAA} logs of remote XrootD accesses; EOS \cite{EOS} logs of file accesses on CERN EOS system; CMSSW \cite{CMSSW} logs capturing information about CMS software framework, and CRAB \cite{CRAB} logs representing user analysis jobs. All of these data are accessible on HDFS.

Each data-stream contains various metrics associated with a job in question, e.g. name of the file used  during a job, user Distinguished Name (DN), processing time of the job, and similar. In order to get insight into data present in these data streams they should be properly cleaned, sorted, filtered, and joined with data from other sources such as DBS and PhEDEx DBs which provide information about dataset, block and site names. The latter two DBs are quite large, their current size is at the level of a few hundred GBytes each. In order to answer simple questions about users accessing a particular dataset on a certain site(s), it was required to join various attributes from multiple streams along with DB tables from DBS and PhEDEx Oracle DBs.
We found that processing time to join and aggregate data across DBS/PhEDEx DBs and data-streams is impossible to achieve in a reasonable amount of time using a vertical approaches based on RDBMS solutions\footnote{A typical job to aggregate information from Oracle DBs and a single data-stream required many hours to complete using materialized view across all data.}.
At the same time a simple dump of DBS and PhEDEx DBs into HDFS and performing a regular Spark job to join desired data-streams with them allows quickly (on a scale of minutes) to get insight into data we were looking for.
For instance, a daily snapshot consists of about 130M log records which were joined at a file level with DBS and PhEDEx DBs, grouped and aggregated to roughly a few thousand records in under an hour of Spark processing time. We measured that monthly stats can be obtained in about 12 hours.
This allowed us to build a series of dashboards to look at data utilization in the CMS experiment.

\subsection{Data Analytics and Monitoring}\label{DataAnalytics}

There are different approaches to data monitoring. On one hand, someone can delegate pushing information from individual GRID sites and applications into the central monitoring system. On the other, such information can be obtained from an individual logs stored in a common place (e.g. central HDFS) via external jobs. The former approach requires proper instrumentation of monitoring metrics and coordination of data injection from all participating data providers. The latter relies on significant processing power to collect and aggregate information in a reasonable time.

\begin{figure}[ht]
\begin{center}
\includegraphics[width=6.3in]{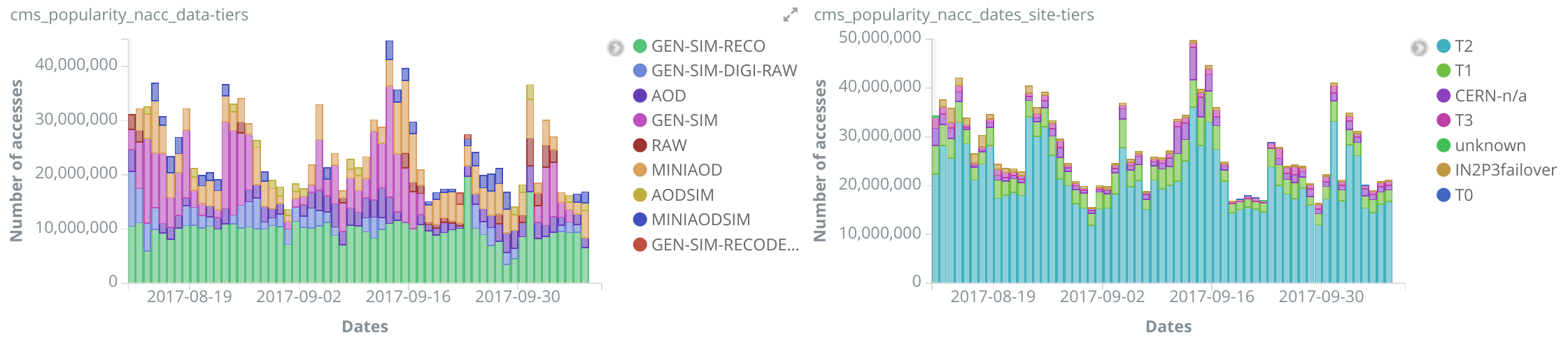}
\caption{CMS popularity dashboard based on processing four CMS data streams: AAA, EOS, CMSSW and CRAB. The displayed part of the dashboard represents the number of accesses vs data-tiers and site-tiers as time series plots.}
\label{fig:fig-dashboard}
\end{center}
\end{figure}

The ability to efficiently process large data sets on HDFS via Spark  jobs opens up a possibility to build various dashboards based on user access metrics extracted from various data-streams. For that purpose we used CERN MONIT system \cite{CERN-MONIT} based on ElasticSearch and Kibana tools. The data processed by Spark jobs were injected as JSON documents into CERN MONIT system via Apache ActiveMQ Stomp \cite{AMQ} protocol. We achieved a reduction factor of 5000 coming from meta-data records (logs) on HDFS to aggregated records pushed into CERN MONIT system.
Fig. \ref{fig:fig-dashboard} shows one part of the CMS popularity dashboard we built.

Even though various dashboard plots were useful to get a broader view on daily activities of CMS users, we were also able to address specific questions raised by various users. For example, how much data resides on all T2 sites with detailed breakdown of number of events and replica sizes of common used data-tiers.

While looking at the current status of resource utilization, which is a very important component of our daily operations, we oversee that it can be further enhanced by predicting user behavior and adjusting our resources accordingly.

\section{Modeling CMS Popularity} \label{Popularity}

Usage of Machine Learning in HEP was mostly used in physics analysis such as discrimination of signal versus background, clustering algorithms, object reconstructions, or similar tasks.
Recently, ML ideas started penetrating into other domains of HEP such as helping physicists better understand computing resources. For instance, ML modeling has been discussed in the scope of dataset placement optimization, reduction of transfer latencies as well as  in network-aware applications for identifying anomalies in network traffic, predicting congestions and WAN path optimization.
Here we present results build on top of previous studies for predicting dataset popularity in context of better data placement strategies. We extend these results by demonstrating how to predict dataset popularity using larger datasets and use them to build strategies for better data placement at participating sites. In fact, DS placement at most WLCG experiments is based on DS popularity which should be considered to make an optimal choice of replication to maximize data availability for processing and analysis needs. As shown in \cite{CMSDataPopularity, ParCo}, CMS dataset popularity can be successfully predicted using ML approach.

\begin{figure}[ht]
\begin{center}
\includegraphics[width=4.5in]{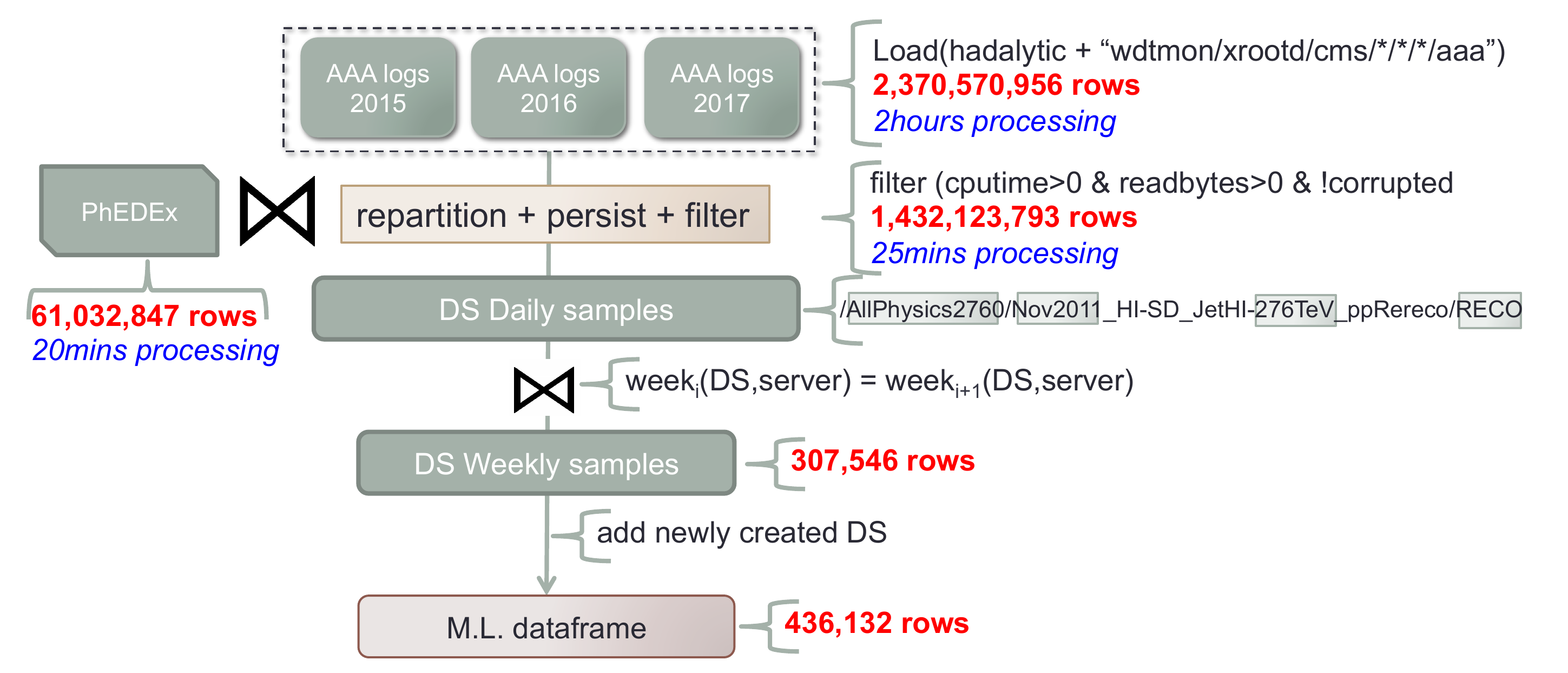}
\caption{An example of reduction pipeline of Spark components from file AAA access logs level to a dataframe suitable for Machine Learning input.}
\label{fig:fig-dataframe}
\end{center}
\end{figure}

In this work, the generation of ML dataframes involved several processing steps on Spark and it was achieved through a Scala reduction pipeline shown in Fig. \ref{fig:fig-dataframe}. It highlights the number of samples and the computing time at each step of the reduction chain. In the shown example, we start from two billion AAA file access logs, merge them with PhEDEx DB to extract dataset and site information and end-up with half a million records suited for ML studies.

\begin{figure}[ht]
\begin{minipage}{19pc}
\includegraphics[width=18.5pc]{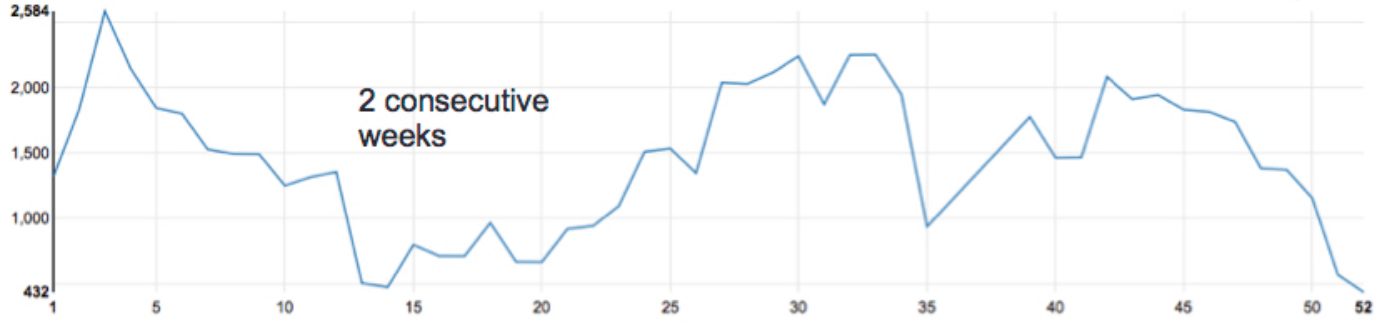}
%\caption{\label{fig:fig-timespan-weeks}DS accesses on a 3 weeks time interval are ~50\% lower than across 2 weeks.}
\end{minipage}\hspace{0.5pc}%
\begin{minipage}{19pc}
\includegraphics[width=18.5pc]{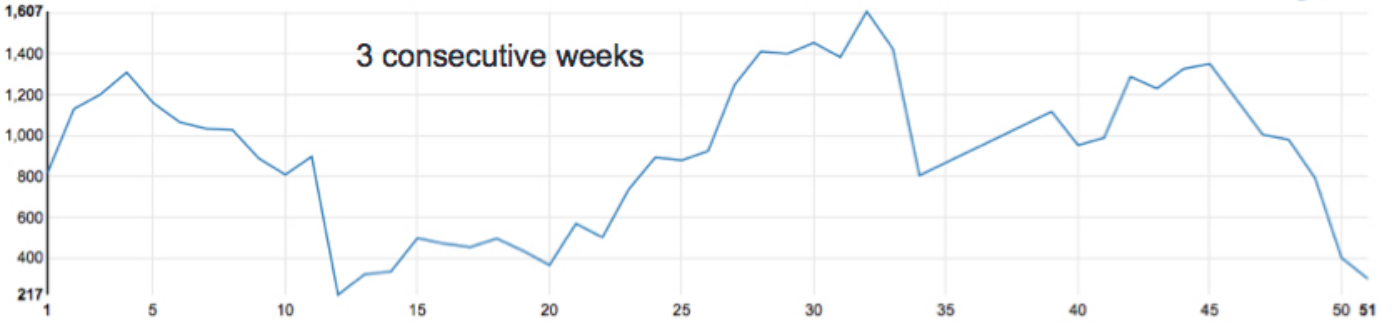}
%\caption{\label{fig:fig-timespan-totweeks}DS have relatively short access pattern.}
\end{minipage}
\caption{\label{fig:DS-popularity}Dataset access patterns during 2 and 3 week period of time.}
\end{figure}

The input dataframe for popularity predictions consisted of a set of attributes and a popularity label. This label is computed on the basis of the usage metric designed as popularity feature. When usage metric in some $time\_period_{i+n}$ is higher than a given threshold, the samples corresponding to $time\_period_{i}$ is marked as popular. 
Our choice for timespan $n$ was chosen to be one week. This was justified by empirical observation. When the temporal window in the training set was increased to 3 weeks the number of positive samples decreases remarkably and prediction accuracy decreased. In fact, Fig. \ref{fig:DS-popularity} shows that the number of DS accesses over 3 weeks time frame is about $50\%$ lower than across 2 weeks, which explains that most DS have a time-limited access pattern.

%In fact, Fig. \ref{fig:fig-timespan-weeks} shows that the number of DS accesses over a 3 weeks time frame is about $50\%$ lower than across 2 weeks. Similarly, Fig. \ref{fig:fig-timespan-totweeks} shows the total number of weeks each DS is accessed and the number of consecutive weeks each DS is accessed. 
%From the two distributions it is clear that most DS have a relatively short access pattern.

%Raw data is available starting from the beginning of 2015. For example, over 2 billions of AAA XrootD federation file access logs are organized in annual groups and concatenated for subsequent repartition across the Spark cluster nodes. 
%At the end of a filtering step, where only traces that are well formatted and contain non-zero reads, the number of samples is halved. These samples are joined with the entries of the PhEDEx file-block location in order to associate each file name the corresponding DS name. The intermediate outcome consists of a set of entries containing either DS strings extracted from its namespace as well as static numeric attributes and usage features. At this stage the entries are aggregated on a daily basis, whereas ML dataframe should be grouped on a weekly basis and should contain the popularity label. This is accomplished performing a self-join on the DS and server names, which brings to the final output suitable for ML algorithms and containing nearly 450k samples.

After applying a cut on number accesses to a dataset, we converted used dataset into classification problem which later was trained on Spark via various classifiers available in Spark MLLib. We achieve the same level of accuracy as reported in \cite{CMSDataPopularity}.

The outcomes from the classifiers are used to implement a data caching policy based on popularity predictions. The datasets are stored at various CMS Grid sites to minimize the network traffic and spread the load for analysis jobs. The current site placement policy is based on assumptions on the expected needs for datasets, mostly using the data-tier. The use of storage, at least for small-to-medium sites, is based on Least Recently Used (LRU) caching systems in order to select which elements should be removed when a site reaches its capacity. Instead, we design a different caching policy that relies on knowledge of DS popularity in the next week and keeps DS in the cache if it is still popular. If a given DS is selected for eviction but is marked as popular by the prediction model, then it is maintained inside the cache.

%\begin{table}[ht]
%  \begin{varwidth}[b]{0.6\linewidth}
%    \centering
%    \begin{tabular}{ l r r }
%      \toprule
%      \textbf{Site} & \textbf{hep.wisc.edu} \\
%      \midrule
%      \#DS & 16,131 \\
% 	\#Popular DS & 4,278 \\
% 	\%Popular DS & 26.52\% \\
% 	\#Reads & 77,975,147 \\
% 	Max weeks/DS & 51 \\
% 	Compulsory Misses & 5,817 \\
% 	Max Hit Rate & 0.64 \\
%      \bottomrule
%    \end{tabular}
%    \caption{Application of PPC to one of the CMS site}
%    \label{table:cache}
%  \end{varwidth}%
%  \hfill
%  \begin{minipage}[b]{0.4\linewidth}
%    \centering
%    \includegraphics[width=80mm]{fig-cache}
%    \captionof{figure}{Hit rates comparison between PPC, LRU and %SDC}
%    \label{fig:fig-cache}
%  \end{minipage}
%\end{table}

%\begin{figure}[ht]
%\begin{center}
%\includegraphics[width=4.5in]{hit-rate-table.pdf}
%\caption{Hit rates comparison between Popularity Prediction Caching (PPC), Least-Recently Used (LRU) and Static-Dynamic Cache (SDC) strategies (left plot) along with breakdown of PPC strategy (right table) for one of the CMS sites.}
%\label{fig:hit-rate-table}
%\end{center}
%\end{figure}

%Fig. \ref{fig:hit-rate-table} shows hit rates comparison between three caching strategies: PPC, LRU, and Static-Dynamic Cache (SDC) \cite{SDC}. 

Preliminary results are detailed in \cite{ParCo}, with the main difference w.r.t LRU observed at lower caching size threshold. We always assume a cold start at each site which generates a number of compulsory cache misses corresponding to the first reference to each distinct DS. 
%Site statistics are detailed in the table to the right which shows the number of total datasets, the percentage of popular ones, the number of compulsory misses and the convergence hit rate.
Our caching strategy outperforms LRU by up to 20\% when the cache size is limited. This makes it very effective in production sites with limited storage or bandwidth. It paves the way for further comparison with others caching strategies and for research aiming at enabling smart policies for dataset replica management based on demonstrated ML approach.

\section[-1pt]{Conclusions and Outlook}

Spark has proven to be a very solid framework for data analytics and ML. It is extremely effective for crunching large datasets available on HDFS and extracting aggregation metrics from underlying meta-data. It can be integrated into HEP workflow via PySpark APIs or stand-along application written in Scala or Python. We discussed possible pitfalls programmers should be considered using either of the programming languages.

We demonstrated that Spark framework represents a valuable alternative to RDBMS solutions for analytics needs. It allows to (re-)process and aggregate various metrics from multiple data-streams efficiently in distributed nature. We were able to access logs from different data streams including AAA, EOS, CMSSW and CRAB and join them with the HDFS dumps of two largest CMS databases, DBS and PhEDEx. As a result the data can be easily aggregated into set of dashboards based on CERN MONIT system and be used for various monitoring tasks within collaboration.

We also implemented a scalable pipeline of Scala components to accomplish mission-critical data reduction and ML processing from billion of records available at CERN HDFS system and build ML models that can be used towards data-driven approach in CMS computing infrastructure. For example, dataset popularity predictions can play a significant role in intelligent data placement at GRID sites both for newly created datasets and holding dataset samples frequently accessible by CMS physicists. A clever caching strategies based on ML predictions can lead to cost-effective data placement across GRID sites and better utilization of available computing resources.

In conclusion, CMS considers the Spark platform as a potentially crucial component in the ecosystem that would enable LHC experiments to attack Big Data problems in the long run.

%We strongly feel that Spark platform will place a significant role in next round of LHC experiments to attack Big Data problems.
%In conclusion, CMS considers the Spark platform - would it be Scala or PySpark - as a potentially crucial component in the ecosystem of enabling technology that would enable LHC experiments to attack Big Data problems in the long run.

\end{document}